\documentclass[10pt,journal,compsoc]{IEEEtran}

\ifCLASSOPTIONcompsoc
   \usepackage[nocompress]{cite}
\else
    \usepackage{cite}
\fi

\ifCLASSINFOpdf
 
\else
 
\fi

\usepackage{graphicx}
\usepackage{xcolor}
\usepackage{tabularx}

\usepackage{authblk}

\begin{document}

\title{Web3 Challenges and Opportunities for the Market}

\author{\IEEEauthorblockN{Dan Sheridan\IEEEauthorrefmark{2},  James Harris\IEEEauthorrefmark{2}, Frank Wear\IEEEauthorrefmark{2},Jerry Cowell Jr\IEEEauthorrefmark{2}, Easton Wong\IEEEauthorrefmark{2}, Abbas Yazdinejad\IEEEauthorrefmark{3}
		}
	\IEEEauthorblockA{
	 \\
					\IEEEauthorrefmark{2}College of Computing and Software Engineering, Kennesaw State University, GA, USA\\  hsherid2@students.kennesaw.edu; Ewong6@students.kennesaw.edu ; jharr651@students.kennesaw.edu; fwear@students.kennesaw.edu; jcowellj@students.kennesaw.edu  \\
												\IEEEauthorrefmark{3}Cyber Science Lab, School of Computer Science, University of Guelph,
		Ontario, Canada  \\
ayazdine@uoguelph.ca\\
							}}




\IEEEtitleabstractindextext{%
\begin{abstract}
\textcolor{black}{The inability of a computer to think has been a limiter in its usefulness and a point of reassurance for humanity since the first computers were created. The semantic web is the first step toward removing that barrier, enabling computers to operate based on conceptual understanding, and AI and ML are the second. Both semantic knowledge and the ability to learn are fundamental to web3, as are blockchain, decentralization, transactional transparency, and ownership. 
Web3 is the next generational step in the information age, where the web evolves into a more digestible medium for users and machines to browse knowledge. The slow introduction of Web3 across the global software ecosystem will impact the people who enable the current iteration. This evolution of the internet space will expand the way knowledge is shared, consumed, and owned, which will lessen the requirement for a global standard and allow data to interact efficiently, no matter the construction of the knowledge. The heart of this paper understands the:
1) Enablement of Web3 across the digital ecosystem.
2) What a Web3 developer will look like.
3) How this alteration will evolve the market around software and knowledge in general.
 }
\end{abstract}

}

\maketitle

\IEEEdisplaynontitleabstractindextext

\IEEEpeerreviewmaketitle

\section{Introduction}
Decentralization, blockchain technology, and token-based economics are all concepts incorporated into Web3, a new iteration of the World Wide Web \cite{ww}. The origin of Web3 can be traced to Tim Berners-Lee \cite{1}, the founder of the World Wide Web and the first person to summarize and discuss the idea of Web3, which he originally referred to as the "Semantic Web". However, Web3 is currently defined by a set of principles focused on decentralization, user ownership of data, and cryptocurrency \cite{3}. Berners-Lee identifies the initial challenge of creating a system that allows knowledge sharing with no central governance or forced commonalities that make sharing inherently tricky. This idea was partially delivered by introducing search engines and globally accepted guidelines on structuring data so that it can be indexed \cite{2}, which helped organize data and make it more accessible. This combination of "do-gooding" developers and growing search engines led to enormous amounts of shared data which is now accessible to a wide audience \cite{2}.

Web3, or the semantic web, has evolved from the idea of growing the Internet out of this inherent paradox of an encyclopedia without a set organization to creator or user ownership of the knowledge and assets that collectively comprise the Internet. The solution, as identified by Berners-Lee, is the ability to pass data along with the rules and logic that govern it from source to source while attempting to lose as little information as possible \cite{2}. The implementation of this solution is now contained in the public blockchains that are the foundation of cryptocurrencies and smart contracts \cite{a5}. This in theory, resolves the limitation of needing developers to create knowledge in good faith to be able to be shared and indexed because of the ability to pass logic across the Internet. Berners-Lee also highlights that the benefits of this idea are not waiting for an astronomical breakthrough in technology where "future software agents…can navigate the wealth of its rich representations." Instead, it can be gradually realized as systems provide more detailed information \cite{2}, and the community makes a conscious effort to adopt those core principles of knowledge sharing enablement and user ownership.
With the connection between systems through smart contracts, the benefits of a Web3 world can begin to be realized as greater change comes over time. This fundamental move to Web3 is not synonymous with a technology upgrade but rather aligned with a growing Internet culture focused on data sharing and ownership that can enable computers to process this data better for the benefit of the wider community. As Berners-Lee puts it, "think of the web today as turning all the documents in the world into one big book, then think [Web3] as turning all the data into one big database, or one big mathematical formula. Web3 is not a separate web; it is a new form of information adding one more dimension to the diversity of the one web we have today" \cite{2}. Simply put, Web1 was read-only. Web2 was read-write. Web3 is read-write-own \cite{4}.

\section{Foundation of Web3 and Enablement}\label{ExSurv}
Web3, as mentioned by Berners-Lee, is not going to hinge on a technological breakthrough, although there is evidence already of creative libraries that enable Web3 ideas. Web3 is an ideological shift in how the Internet of data is constructed that can be enabled with existing technology to produce benefits today. The enablement of Web3 will depend on the transition to Web3 libraries and concepts by knowledge and data creators. The foundational items that currently comprise the Web3 framework are blockchain networks (decentralized \cite{a6} but connected nodes), Web3 libraries, emerging specialized languages (Solidity), identity stores (wallets), smart contracts, and specialized service providers \cite{5}. 

\subsection{Blockchains (Bitcoin, Ethereum, etc.)}
The public blockchains are the core pipelines for all transactions and interactions that occur within Web3 applications. They are the public ledgers of record, decentralized and immutable \cite{a7}. A global permissionless database that provides the ability to track the ownership chain of any asset or piece of data that exists on the blockchain at any time by anyone. While made popular by cryptocurrencies, their utility extends to any form of data. This is about Facebook or any other social network moving your profile to a blockchain where you have complete control to grant, revoke, or even sell your data rather than giving it away for the utility of using the social network. In Web2, the network controlled and held all the value. The principles of Web3 dictate that the data holds all the value, not the network. Web3 is an ideological shift for the Internet, just as Web2 was for Web1.
As users interact across Web3 and its associated frameworks, there is a functional need to house the necessary transactions and interactions. The Web3 components which provide this service and enable blockchains to function are known as nodes. Without nodes, the applications cannot communicate; thus, the decentralization of applications becomes voided. Nodes facilitate the tracking of data and are the multiple storehouses that exist to fight against data loss threats \cite{5} and have additional copies of the interaction. This also makes transactions verifiable and immutable because this would be the corroborating source for other Web3 applications.
\subsection{Web3 Libraries}
The Web3 libraries provide a set of application programming interfaces to bridge blockchains and smart contracts, enabling the creation of a new class of applications, commonly known as Decentralized Applications (aka DApps). By their very definition, Web3 applications utilize a blockchain. Ethereum is the most common blockchain used by DApps as Ethereum was created to support application development and has a governance model specifically crafted to be developer friendly. Web3.js, Next.js, Ether.js, and Truffle Suite \cite{5} are commonly used Web3 application development libraries, which are all JavaScript based. Cloudflare provides an example of this interaction here, which hosts an open-source NFT project to explain Web3 and blockchain \cite{6}.

   \begin{figure}
    \centering
    \includegraphics[width=10cm]{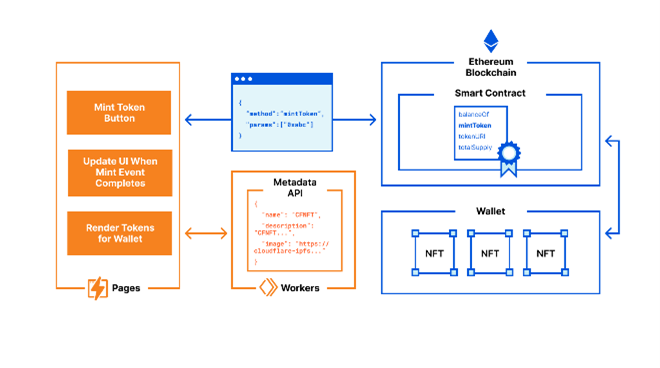}
    \caption{	System targets from a user-centric explainable AI framework \cite{5}}
    \label{fig:life}
\end{figure}
\subsection{Identity Stores (Wallets)}
	With ownership being at the heart of the Web3 principles, immutable identities are a requirement. Wallets are applications that store and protect the identity used for interacting with blockchain and facilitate the literal transaction. This wallet is the “final authority of your data” and can be any entity \cite{6}. Transactions are totaled here, and all the interactions with knowledge will also be tracked here. 

\subsection{Smart Contracts}
	A smart contract \cite{SINGH2020101654} is akin to what Berners-Lee was talking about when passing the rules along with the knowledge. Smart contracts act as a conditional mechanism to make something happen that is fully transparent and separate from any intermediary to fulfill a transaction. Smart contracts rely on nodes to validate the data of a wallet account and pass in the parameters to provide the output of a smart contract condition. Essentially, smart contracts are codes that reflect the logic of the transaction two parties are executing. Every interaction with a smart contract is then recorded in a ‘block’ on the ‘chain’ by the node executing the contract \cite{7}. These would act as the access points of knowledge sharing permission or a point of passing the universal logic.
	These concepts have contributed to the early landscape of Web3 and are critical cogs to a functioning network of decentralized data points, knowledge sharing, and registered owners. 

\subsection{Virtual Reality Systems}
Virtual reality has evolved beyond games today, and people use it for art, tourism, and industrial purposes \cite{v}. Ticketing or fees must be paid for this purpose, and a blockchain is an excellent solution for completely secure payments \cite{v,g}. Blockchains provide a distributed cryptographic ledger that cannot be trusted. There is no need for a trusted authority to verify a party's interaction \cite{a8,a9}. Depending on how users log in and their access rights, there are different types of blockchain, including private, public, and consortium. Generally, blockchain-based virtual reality platforms use a blockchain for supporting Web3 so that anyone can easily participate. Virtual reality in the blockchain process involves hashing all the information the user provides, then encrypting the private key on the user's device and generating a digital signature. On a peer-to-peer (P2P) network, transaction data is sent to peers along with a digital signature. Using the public key, network members decrypt the device by comparing its hash against the hash of the transaction data.  Due to the widespread use of Web3 technology, concerns have been raised about information theft, hacking, dissemination, and copying. Consequently, many companies are using virtual reality, because they are unable to limit supply. To support virtual reality in the middle of Web3, a blockchain is an ideal solution for creating decentralization, security, and transparency \cite{a10,a11}. The data recorded in the virtual reality system cannot be altered, and its use can be easily traced. As a result of utilizing blockchain technology, security is increased, and the user experience is improved. Virtual reality security and trust can be improved by utilizing blockchain.

\section{A Web3 Developer}
	Web3 as a concept has created a market for developers with skillsets that can enable and manage the core functions of knowledge and data-sharing-centric network. With Web3 being tied heavily to the current cryptocurrency infrastructure that is aligned with the core principles of Web3, there has been a large move of users contributing to the most significant public GitHub repositories involving the Web3 development stack, according to Richard MacManus at the New Stack \cite{8} due to the interest stirred by the recent popularity of cryptocurrency. This population of around 18,000 developers is likely an underestimate of the active base, as there is no information on the true amount of private proprietary Web3 developers. Consensys, the makers of popular Web3 development tools and services, reported 350,000 active developers using its Infura blockchain development platform in November 2021 \cite{9}. By April 2022 – the time of the creation of this paper – this number had grown to 430,000 active developers \cite{10}. According to the New Stack, this figure is a “drop in the bucket,” with the likelihood of this population growing more in the coming years as this market has the potential to rapidly expand if the market for decentralized and interconnected applications becomes a wider norm for software and technology companies.
	In this market, a Web3 developer does not look vastly different from other developers, with the difference being experienced in the core concepts of blockchain, digital wallets, smart contracts, and the Web3 libraries mostly written in JavaScript and Web3 service providers. Familiarity with the core libraries is important in creating the core functions of a Web3 app which is no different from other fields of software development and relies on knowledge of some of the most popular languages to create this interaction, namely JavaScript \cite{5}. The one exception to this assertion is in the area of smart contracts, which are usually written in a specialized language in order to optimize performance. Some of these languages are Solidity, Rust, Yul, Vyper, and JavaScript. This area is also not restricted in any sense, with open-source libraries available to whoever wants to interact and use them. Web3 development looks to involve more of a fundamental mindset shift in how applications and functions need to interact and the creation of readable scripts that other applications can tap into as portable nodes rather than querying a central server, very similar to querying cloud services. The ability to communicate with the recording system is essential in being part of this Web3 environment because all interactions across the environment rely on an audit log with specific conditions that need to be passed, sometimes referred to as a ledger – the blockchain. Experience with data security is also a key trait for Web3 developers based on the general infrastructure that Web3 is founded on, passing data around transactions through ledgers and conditional function codes known as smart contracts. This interaction is different from most other central-based systems, where logging is always a reference point back to internal systems through various inputs rather than a wider network of checks that are highly portable and accessible via a blockchain network. Security on a blockchain is enforced at every entry and execution point on the chain. If you do not meet the security requirements of the blockchain, your application will not be able to execute smart contracts or write to the blockchain. The governing body of the blockchain sets the security standards, or the DAO – Decentralized Autonomous Organization – their published bylaws.
	Fundamentally, Web3 developers look very much like other software developers today but work with specialized concepts and tools specific to blockchain technologies. Demand will continue to grow for Web3 developers. The population of around 430,000 will also enable these first-comers to make potentially ground-breaking contributions to the space and push Web3 forward. The market will continue to grow for some time, and the developers looking for new challenges and the likes of the ’do-gooding’ developers that helped push Web2 are most likely going to join in the coming years. The continual fascination with blockchain as a technology and its applications will also be a part of this growing market with estimates as high 34.1\% CAGR (compounded annual growth rates) over the next 10 years \cite{11}. With blockchain being closely tied to Web3 as an enabler, this will create more developers capable of pushing the Web3 principles in other areas outside the core areas right now of banking, cybersecurity, and cryptocurrency. Developers wishing to explore a shift in their careers can find many helpful free resources, from job boards \cite{12} to guides to learning Web3 technologies \cite{13,14}. 
	\section{Web3 Impact}
\begin{figure}
    \centering
    \includegraphics[width=8cm]{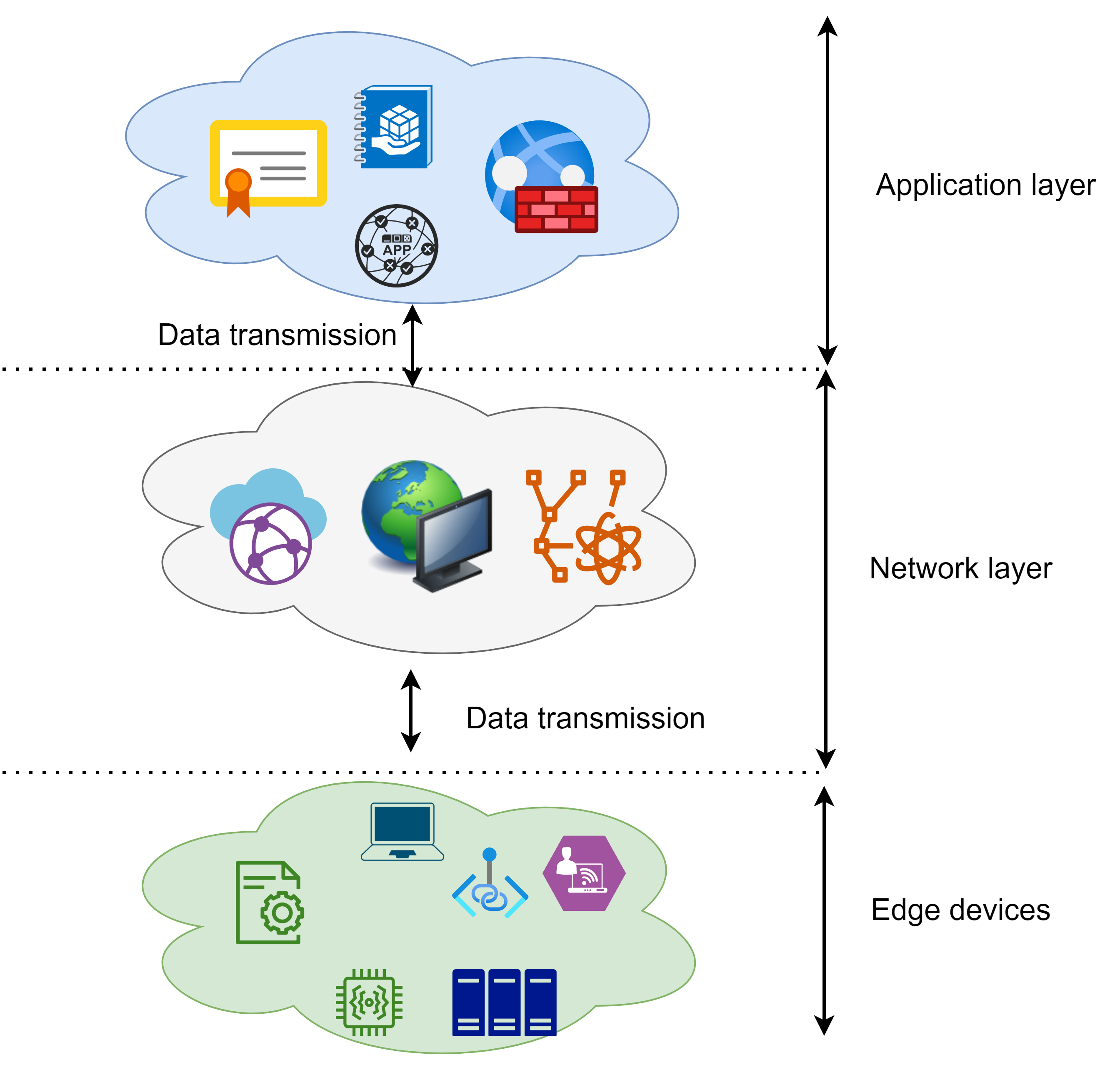}
    \caption{The place of web in the network architecture }
    \label{2}
\end{figure}

	In order to understand Web3, we must first understand the place of the web in the network architecture, as shown in Fig \ref{2} In this way, real-world applications can help us better understand web functionality.

	Web3 aims to be an evolution into an open, permissionless, decentralized environment for the Internet that will have widened the scope of knowledge and data accessibility and fundamentally shifted the economics upon adoption \cite{15}. Whether this adoption is contingent on technology or ideology, these principles will hold true on the overall impact in both the market and the internet ecosystem. The core impact will be on how computers will interact with other computers for the benefit of users and who owns the outcome. This will create smarter searches and better application of smart technology to put the ownership of the outcomes in the hands of the users and creators rather than the intermediaries of Web2.
	Web3 will be structured as open-source code, not reliant on the big middlemen that enable accessing of content today, such as Google, Microsoft, and Facebook as examples \cite{15}. That is until Big Tech creates their own blockchains, makes them public, and incentivizes developers with airdrops backed by the company's stock to use their tools and services, making the developers owners of the company. The development of Web3 has become transparent to allow software developers to utilize the growing library of publicly accessible Web3 packages and libraries to build the ecosystem as soon as information becomes available, accelerating the velocity of evolution.
	This availability of the infrastructure will enable users to create this Web3 ecosystem. A core piece of this idea is around permissionless technology where interactions are not reliant on trusted third parties to connect users, whether through a search engine or transaction machine \cite{15}. The goal will be to make this feasible by "blockchain-like" infrastructure, if not the same libraries that blockchains use today. The structure acts as its intermediaries, so peer computers can cut out the requirement of an intermediary by using universal logic and conditions to have transferable data ledgers to track transactions and share knowledge based on rules within the node through smart contracts. This helps to protect against the concern of data theft by the current infrastructure of Web2 that enables intermediaries to capture your data and use it for their purposes, often monetizing it in ways the creators of the data might not agree with. Allowing the rules and conditions to be transferable and available without going through an intermediary can also increase execution speed, transparency, and equitable ownership.
	The reliance on large sole sources in the form of data warehouses that maintain knowledge is most likely to be disrupted by the principles of knowledge within Web3. Adopting Web3 as the infrastructure creates more areas where data is stored, which mitigate major data-loss threats, keeps data behind rules not maintained by one guardian, and keeps data from being hidden and controlled by middlemen that disperse the data not necessarily to the benefit of the user base \cite{15}. Decentralization of data storage is a core feature of Web3 that will allow users to create more content and follow a universal logic to share information rather than conforming to an intermediary that acts as a gatekeeper that pushes a framework requirement to best share out the work \cite{15}. Lessening the effort required to create new content accessible can make the whole process of knowledge sharing and value creation a lot faster than it is today through the use of Web3 principles and technology.
\section{Web3 Risks}
	With the emergence of this new Web3 ecosystem which is projected to have a large swath of influence, there are some inherent risks based on the core infrastructure of Web3 and blockchain technology as a whole \cite{16}.
\subsection{Lack of Regulation and Oversight}
	Today, blockchain technology interfaces within the Web3 landscape are essentially unregulated, with a general lack of understanding by most regulators. There are no written laws or advisory boards overseeing how this ecosystem operates, which poses a risk of bad actors and bad faith interactions occurring. This open space of operation has regulatory boards struggling to define even the structure of what are ’entities’ within this operation, such as how even to define what a “blockchain-enabled organization” is in comparison to other companies \cite{17}. Blockchains are operated by Decentralized Autonomous Organizations (DAO’s) with varying and non-standard governance models by common business standards \cite{a12}. DAOs are organizations represented by rules encoded as a transparent computer program, controlled by the organization members, and not influenced by a central government \cite{18}. 
	Transactions pose another risk based purely on the global scope of how Web3 can interact, with a central point not existing in any of these ecosystems. For knowledge sharing of particular materials, the jurisdiction of regulation is unclear because it can exist in many different environments and tracking a geographic origin might be impossible. Countries, regions, and governing bodies have different regulations for more traditional knowledge sharing that would be disrupted by how Web3 and blockchain operate without a framework to govern this. With services being provided through a Web3 environment, there are also financial implications in how those are served and to what body. It is unreasonable to believe that because there is not one singular regional source that governing bodies will allow transactions to be monitored/taxed without oversight for a long period of time. This would include how bad faith transactions would work because currently, there is no clear operating system to govern and ensure people are behaving appropriately beyond the bylaws of the DAOs. Traditional ‘scams’, insider trading, and various schemes can be adjudicated through justice systems that are not connected to or cover the framework of Web3’s blockchains. This new business model of promoting peer-to-peer transactions now puts the risks with users instead of what used to be managed by central intermediaries, which is an inherent risk of working with blockchain technology \cite{19}.
	For example, a current blockchain/cryptocurrency scheme that exists (which does not directly tie into Web3 and the semantic web but provides an overview of potential risks) is referred to as a crypto pump and dump, which is parallel to insider trading on big events. An organization will promote a cryptocurrency they may have a large stake in to push the value up considerably before cashing out and typically crashing a currency with new entrants bearing the brunt of the losses. Because most governing bodies do not consider cryptocurrency as a security, let alone a commodity or actual currency, there is no legal precedent for doing anything about this practice that is parallel to an illegal activity \cite{20}.
	As Web3 is adopted and operates for a wider audience, regulation boards should move closer to providing guidelines and direction to ensure the safety of consumers and businesses from bad actors \cite{a13}. This will also be a big pillar of credibility to this area of what exists now as the gray area with few legal definitions.
\subsection{Portability of Illegal Items}
Web3, as professed in the previous material, is focused on allowing knowledge to be shared without a required framework where information can be easily queried and passed along with the rules that define it. The ecosystem now exists within a connected framework where each node houses the same universal information that makes traceability hard. This means bad actors can use these properties to create material that is easily accessible by the people who want to view it and hidden from users who do not want to view it through a major component of Web3 as the smart contract. Illegal media could more easily exist in a Web3 framework due to this decentralization that allows creators to be more inclined to release more of this knowledge and data due to the inherent security provided by a decentralized and conditionally accessed piece of knowledge.
	As a concept, the portability of illegal items might make other internet content creators and providers hesitant to begin the full-on shift based on the services they plan to make available, especially with no plan/guideline in place to protect against these questionable practices that, in parallel mediums are considered illegal \cite{a14}. This gap has a market in waiting, so potentially by the time Web3 is truly ready to be embraced, companies and services could be available to help combat this issue with their own version of a secure smart contract and ledger. This could also be in combination with wider efforts of regulators to put in place guidelines on the operation and have a better understanding of this space operates.
	
\subsection{Accuracy of Information}
	Now that, in the context of Web3, content is easily transmissible and not contained to a central source, there lies a risk with the rapid movement of disinformation \cite{21,a13i}. The ability to send and receive information fast via a Web3 framework has issues in highlighting good or bad information. Current state, content creators and providers have the opportunity to monitor this information and correct items because it is sourced in their own central area. Within a Web3 framework, information can pass way faster with no restrictions and no oversight. This issue is prevalent today through other semi-decentralized applications such as WhatsApp, which has had rampant issues in combating disinformation in its own encrypted messaging platform \cite{21}.
	
\subsection{Privacy}
	With security at the forefront as a benefit for the Web3 framework, there is still a vague gap regarding data privacy and how interactions will be saved not only for a user's wallet but also within the network \cite{20}. If an interaction is logged within a ledger and that ledger exists everywhere, what sort of information does the wallet provide to get in, and will users be able to understand those implications? This could mean now intermediaries do not control personal data but potentially their peers. With data being free flowing in this environment, there seems to be larger hurdles to surpass for data privacy, primarily when sources are not centered in regions but exist across networks. A user-focused approach for Web3 empowers users and the internet to be focused on them, but that does not provide them the education on how their information is going to be processed by each unique system, and this especially means regulators will not understand this either\cite{20,a15}. 
	Web2 policies around data privacy do not account for this framework and will need to be updated to account for the new capabilities of the framework to process information and store it universally. This alone will put the full onus on service providers to ensure they store this information correctly, which is not sufficient without regulatory oversight to provide clear direction on what that means. There is danger in allowing companies to govern themselves because ensuring that you are constantly doing the right thing does not always fit with the plan, as we can review countless firm practices that have put consumers and users in harm's way.
	\subsection{Future Roadmap}
Our digital world faces new technologies ( e.g., Software Defined Network (SDN) \cite{t}, Network functions virtualization (NFV) \cite{a1}, Artificial intelligence (AI) \cite{Ekramifard2020}, Internet of things (IoT)  \cite{a2}, and Blockchain \cite{a4}) that directly and indirectly affect human life in education, business, industry, and the military. Combined technologies have also improved construction, energy, and resource efficiency \cite{a1,a2}. Web3 \cite{w} is one of today's hottest, most attractive, and most exciting technologies.

	We anticipate that research on Web3 will move towards quantum, IoT, AI, and secure cloud-assisted areas. The reason behind this anticipation is the existence of trends toward the application of security, Quantum, AI, IoT, and secure cloud-assisted environments are interesting areas for academic and industry sections \cite{a17}. Many companies are investing in Web3 applications and building their products to support them.
   \begin{figure*}
    \centering
    \includegraphics[width=18cm]{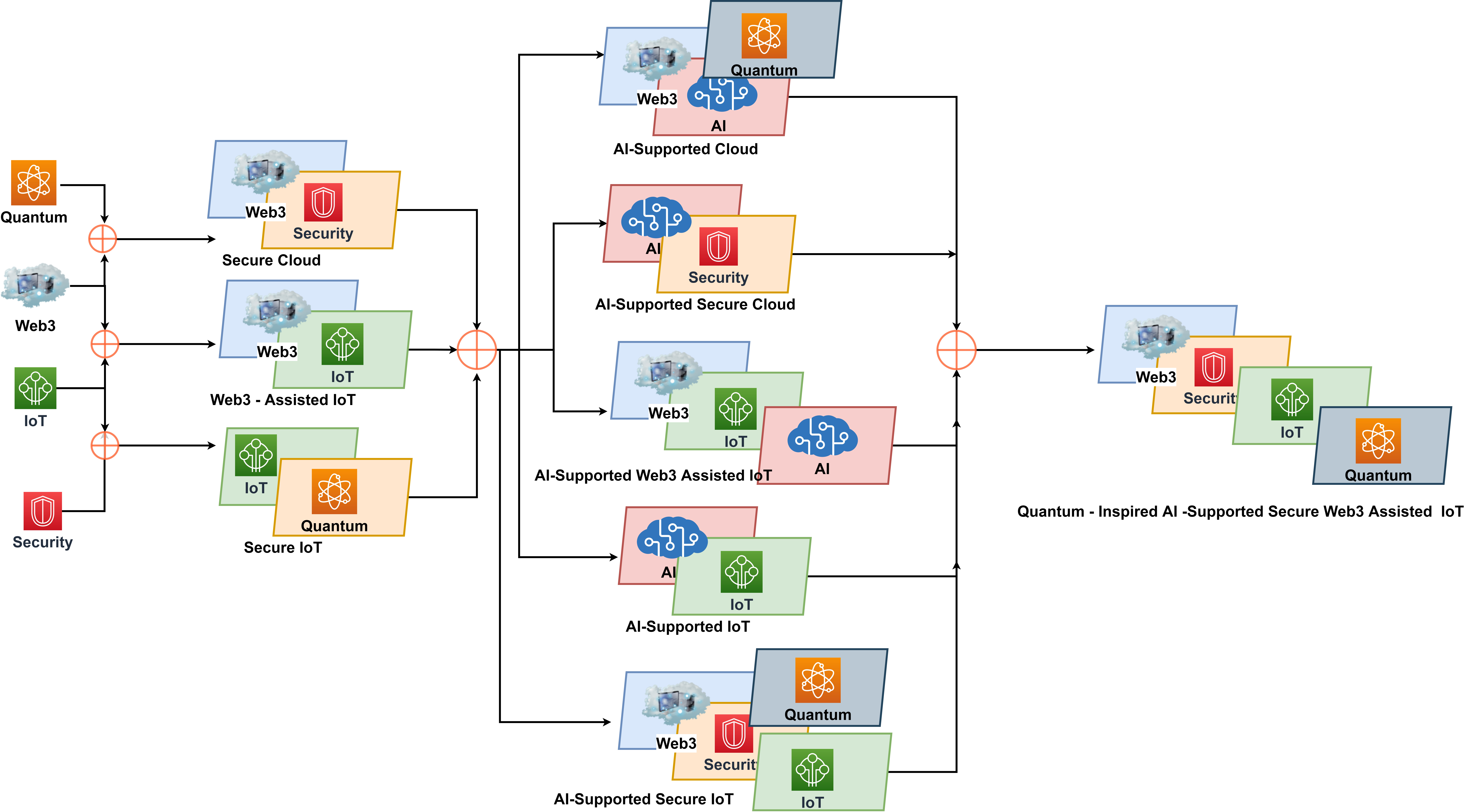}
    \caption{	Combination of Web3 with other technologies }
    \label{3}
\end{figure*}
Moreover, Fig. \ref{3} summarizes the widespread involvement of AI and  IoT, Quantum with Web3, for the future. Indeed, it shows that inspired AI supported Web3 and the ability to Combination of Web3 with other technologies. It can prove these areas will move toward Web3 quickly.
\subsection{Conclusion}
Fast forward 10 years – all developers are Web3 developers of some form. The definition of "full stack" will evolve. Front-end developers will remain focused on the user experience but may have to deal with additional complexities of data coming from blockchains or sending data for smart contract execution. Back-end developers will still deal with server or cloud application-based logic, but the new complexities of blockchain and smart contract dependencies will require new skill development. Blockchain and smart contract layers will be added to the definition of "full stack", creating new areas of specialization and opportunities. With each blockchain transaction being assigned a very specific value in terms of the transaction itself and the value of the data contained in the blocks, security will become a very acute focus and a specialization unto itself. Web3 security will be focused on the security of the blockchain and the smart contracts, the points where the value is created and stored. Security practices will become ever more integrated into the code of the applications, increasing the requirement of developers to obtain new skills and demonstrate secure coding and development practices. The inherent risks posed by the current state of Web3 technologies create opportunities for growth and value creation. Likely, the jobs boom that occurred as the industry began to understand what Web2 meant in the context of Web1 will be dwarfed by the jobs boom Web3 will generate. The fundamental economic shifts in the ideology of Web3 alone will trigger a tectonic shift we have never seen before. This, coupled with the scarcity of resources the industry expects over the next decade, even without considering the impact of Web3, will provide developers and other roles in software engineering a tremendous opportunity to generate value and wealth \cite{22,a21,23}.

\bibliographystyle{IEEEtran}
\bibliography{References}

\end{document}